\documentclass[prl,preprint,floatfix]{revtex4}
\usepackage{graphicx}
\begin{document} 
%{\Large{\bf Non-monotonic enhancement of synchronization with
%      increasing randomness in coupling connections}}\\
\title{Enhancement of spatiotemporal regularity in an optimal window of random coupling}
%\title{Existence of an optimal window of randomness in coupling connections for enhanced synchronization}
\author{Swarup Poria {\footnote{e-mail: swarup$_-$p@yahoo.com}}}
\affiliation{Department of Mathematics, Midnapore College, Midnapore,
  721 101, West Bengal, India} \author{Manish Dev
  Shrimali{\footnote{e-mail: m.shrimali@gmail.com}}}
\affiliation{Department of Physics, Dayanand College, Ajmer, 305 001,
  India} \author{Sudeshna Sinha{\footnote{e-mail:
      sudeshna@imsc.res.in}}} \address{The Institute of Mathematical
  Sciences, Taramani, Chennai 600 113, India}

%\maketitle
\begin{abstract}
%Interestingly the basin of attraction for the synchronized state
%  varies non-monotonically with rewiring probability $p$.

We investigate the spatiotemporal dynamics of a lattice of coupled chaotic maps
%with varying degrees of randomness in coupling connections, with
whose coupling connections are dynamically rewired to random sites
with probability $p$, namely at any instance of time, with probability
$p$ a regular link is switched to a random one.
%  with its links randomly rewired with probability $p$. 
%dynamically switched random rewirings, with the rewiring
%  probability being $p$ ($0\le p \le 1$). 
In a range of weak coupling, where spatiotemporal chaos exists
for regular lattices (i.e. for $p=0$), we find that $p>0$ yields
synchronized periodic orbits. Further we observe that this regularity
occurs over a window of $p$ values, beyond which the basin of
attraction of the synchronized cycle shrinks to zero. Thus we have
evidence of
%pronounced enhancement of synchronization in a window of $p$, i.e. there exists an the existence of 
an optimal range of randomness in coupling
connections, where spatiotemporal regularity is efficiently obtained. This is
in contrast to
%the scenario where regularity is induced only in the limit of small $p$, 
%and it is also distinct from 
the commonly observed monotonic increase of synchronization with
increasing $p$, as seen for instance, in the strong coupling regime of
the very same system.

%Further, under increased interaction range, or
%  with static random rewirings, the system again exhibits the usual
%  monotonic enhancement of synchronization with increasing $p$.

\end{abstract}

\maketitle

%89.75Hc  Networks and genealogical trees 05.65.+b Self-organized systems (see
%also 45.70.-n in classical mechanics of discrete systems) 87.19.Xx
%Diseases 87.23.Cc        Population dynamics and ecological pattern formation
%87.23.Ge        Dynamics of social systems 64.60.-i        General studies of
%phase transitions 02.50.Ey        Stochastic processes 05.45.-a
%Nonlinear dynamics and nonlinear dynamical systems

\section{Introduction}

The dynamics of spatially extended systems has been a focus of intense
research activity in the past two decades. In recent years it has
become evident that modelling large interactive systems by finite
dimensional lattices on one hand, and fully random networks on the
other, is inadequate, as various networks, ranging from collaborations
of scientists to metabolic networks, do not to fit in either paradigm
\cite{RMP_bar}. Some alternate scenarios have been suggested, such as
the small-world network \cite{Watts}. Here one starts with a regular
structure on a lattice, for instance nearest neighbour
interactions. Then each regular link from a site is {\em rewired
  randomly} with probability $p$.  This model is proposed to mimic
real life situations in which non-local connections exist along with
predominantly local connections.

There is much evidence that random nonlocal connections, even in a
small fraction, significantly affect geometrical properties, like
characteristic path length \cite{relev}. However its {\em implications
  for dynamical properties is still unclear and even conflicting}.
%So the first question we will probe here is this: {\em does one see
%  dynamical changes at very low values of $p$}, namely does the
% behavior change as soon as non-local shortcuts are put in place
%\cite{relev}.
While the dynamics of coupled oscillators and coupled maps on regular
lattices has been extensively investigated \cite{cml}, there have been
far fewer studies on the sptiotemporal features of nonlinear elements
on more general network topologies \cite{dyn_net,sinha}.
%\cite{dyn_net,sinha,non_mono}.

Most existing case studies of coupled networks of dynamical elements
indicate that features, such as degree of synchronization, vary {\em
  monotonically} with $p$. That is, it is observed that most
dynamical properties interpolate between the limits of regular and
random connections without in any sense being ``optimal'' or more
pronounced at some intermediate value of $p$.

In this paper however we will provide evidence of a system where there
exists a window of randomness where one obtains special dynamical
features which cannot be anticipated from a simple interpolation
between the regular and random limits. In particular we will show the
pronounced enhancement of spatiotemporal order in the system, in am
intermediate window of rewiring probability $p$. Our observations are
%in contrast to the scenario where regularity is induced only in the
%limit of small $p$ \cite{non_mono}, and is also 
markedly distinct from the commonly observed monotonic dependence of
synchronization properties on $p$, as seen for instance, in the strong
coupling regime of the very same system \cite{sinha}.

\section{Model}

Specifically we consider a one-dimensional ring of coupled logistic
maps. The sites are denoted by integers $i = 1, \dots, N$, where $N$
is the linear size of the lattice. On each site is defined a
continuous state variable denoted by $x_n (i)$, which corresponds to
the physical variable of interest. The evolution of this lattice,
under standard interactions with the nearest neighbours on either
side, in discrete time $n$, is given by
\begin{equation}
%  x_{n+1} (i) = (1 - \epsilon) f(x_n (i)) + \frac{\epsilon}{2K} \sum_{j=-K}^{K} x_n (i+j)
  x_{n+1} (i) = (1 - \epsilon) f(x_n (i)) + \frac{\epsilon}{2} \{ x_n (i+1)+x_n(i-1) \}
\end{equation}
The strength of coupling is given by $\epsilon$. The local on-site map
is chosen to be the fully chaotic logistic map: $f(x) = 4 x (1 - x)$,
as this map has widespread relevance as a prototype of low dimensional
chaos.

Now we will consider the above system with its coupling connections
rewired randomly in varying degrees, and try to determine what
dynamical properties are significantly affected by the way connections
are made between elements. In our study, at every update we will
connect a site with probability $p$ to randomly chosen sites, and with
probability $(1-p)$ to nearest neighbours, as in Eqn.~1.  That is, at
every instant a fraction $p$ of randomly chosen nearest neighbour
links are replaced by random links. The case of $p = 0$ corresponds to
the usual nearest neighbour interaction, while $p = 1$, corresponds to
completely random coupling. Note that the random connections are {\em
  dynamic} here, as the random links are switched around at every
update.  {\em This is in contrast to most studies which consider
  static (quenched) random connections.}

\section{Emergence of Syncrhonized Cycles from Spatiotemporal Chaos}

We will now present numerical evidence that random rewiring has a
pronounced effect on spatiotemporal order. The numerical results here
have been obtained by sampling a large set of random initial
conditions ($\sim 10^4$), and with lattice sizes ranging from $50$ to
$1000$.

Fig.~1 displays the spatiotemporal state of the network, $x_n (i) , i=
1, \dots N$, with respect to coupling strength $\epsilon$, for the
case of regular nearest neighbour interactions (i.e. $p = 0$) and for
the case of random coupling with probability $p = 0.19$ and
$p=0.6$. It is clearly seen that the standard nearest neighbour
coupling does not yield regularity anywhere in the entire coupling
range, while randomly rewiring with small probability $p= 0.15$
creates a window in parameter space where {\em synchronized cycles
  gains stability}.  Note that {\em different periodcities} of the
synchronized cycles are obatined under different coupling strengths,
in the window of synchronization. Interestingly however, when $p$ is
large this window of complete spatiotemporal regularity is lost again
(see Fig.~1c).

\begin{figure}[ht]
\label{fig1}
\begin{center}
\includegraphics[width=0.85\linewidth,height=7cm]{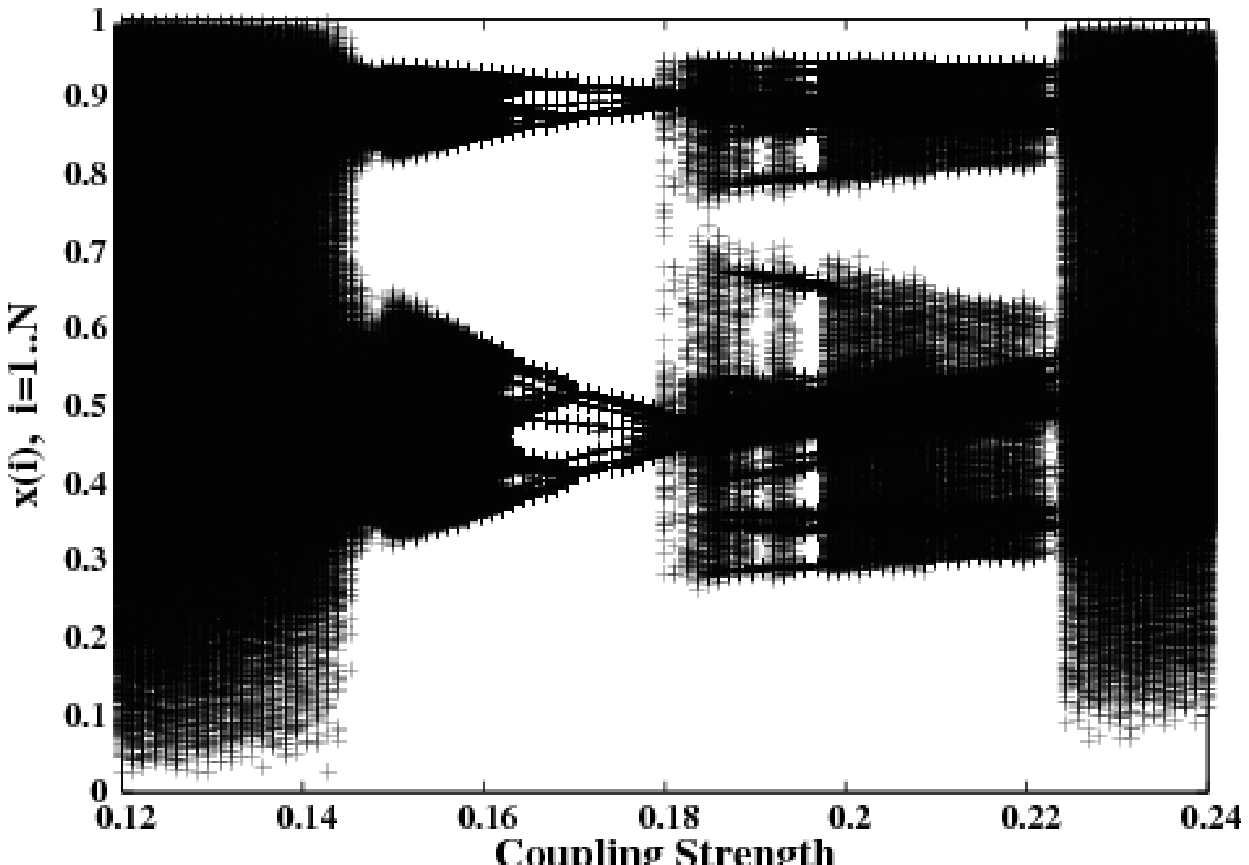} (a)
\includegraphics[width=0.85\linewidth,height=7cm]{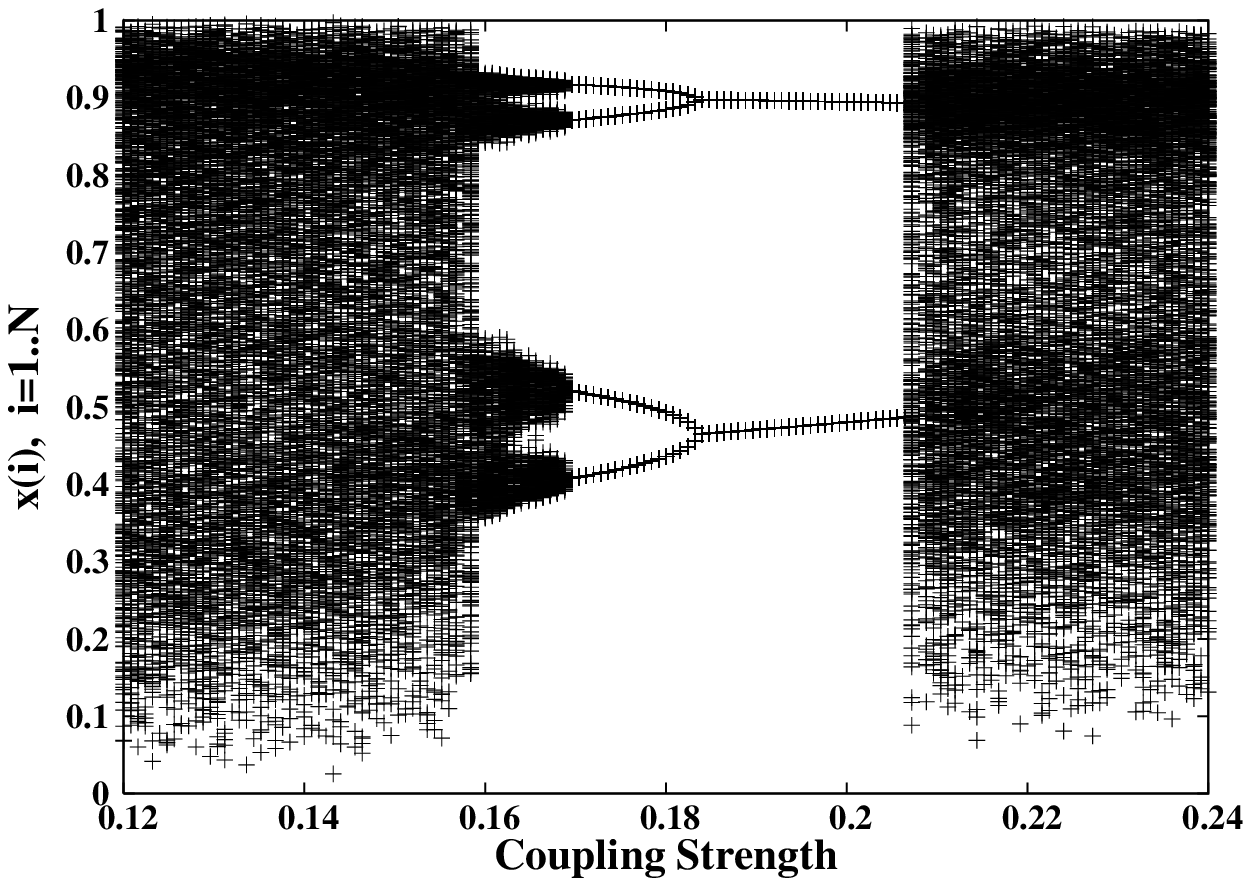} (b)
\includegraphics[width=0.85\linewidth,height=7cm]{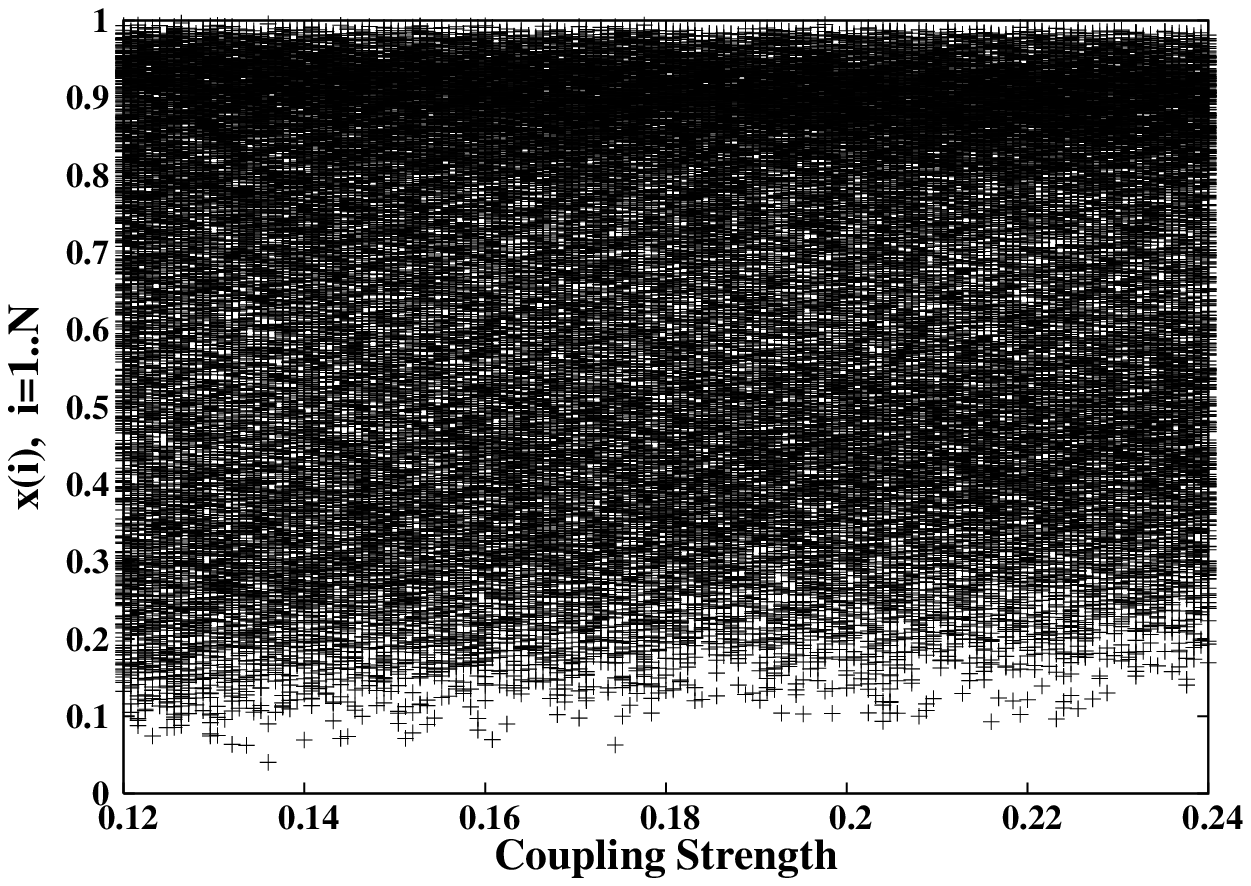} (c)
\end{center}
\caption{Bifurcation diagram showing the state of the lattice $x_n
  (i), i=1, \dots N$ ($N=50$), over $n = 1, \dots 25$, with respect to
  coupling strength $\epsilon$, for (a) $p=0$, (b) $p=0.19$ and (c)
  $p=0.6$. The synchronized $2$ and $4$ cycles are clearly evident in
  (b), but not in (a) or (c).}
\end{figure}

In order to quantify this phenomena, we find the fraction of random
initial conditions that get synchronized (after long transience) This
provides a measure of the size of the basin of attraction of the
synchronized state, denoted by $B$ (see Fig.~2). When $B = 1$ we
obtain a global attractor for the synchronized cycles.

Fig.~3 shows the synchronized basin size $B$ with respect to rewiring
probability $p$, for different coupling strengths. 
%Fig.~3 displays this basin size with respect to coupling strength
%$\epsilon$ for various values of random rewiring probability $p$. 
It is clearly evident from the figure that for the regular lattice
limit of $p=0$, the synchronized basin size is close to zero. As $p >
0$ the basin for the synchronized state increases rapidly. For
instance, for $p=0.01$ in Fig.~3, the synchronized state is the global
attractor of the dynamics, i.e. $B=1$. However, as $p$ exceeds an
optimal window, the basin size decreases rapidly, becoming zero again
for high $p$. Thus there {\em exists a range of $p$ values where
  synchronized cycles gain stability}. This observation is quantified
in Fig.~4 which gives the range of rewiring probability, $R$, over
which $B=1$ for different coupling strengths.

%\begin{figure}[ht]
%\label{fig3}
%\begin{center}
%\includegraphics[width=1\linewidth]{p_0_0.1_0.3.ps}
%\end{center}
%\caption{Size of the basin of attraction of the synchronized state vs
%  coupling strength $\epsilon$ for a system of size $100$ for (a)
%  $p=0$, (b) $p=0.1$ and (c) $p=0.3$. Here transience is $50000$. Note
%  that {\em different periodcities} of the synchronized cycles are
%  obatined under different coupling strengths, in the window of
%  synchronization.}
%\end{figure}

\begin{figure}[ht]
\label{fig3}
\begin{center}
\includegraphics[height=1\linewidth,angle=270]{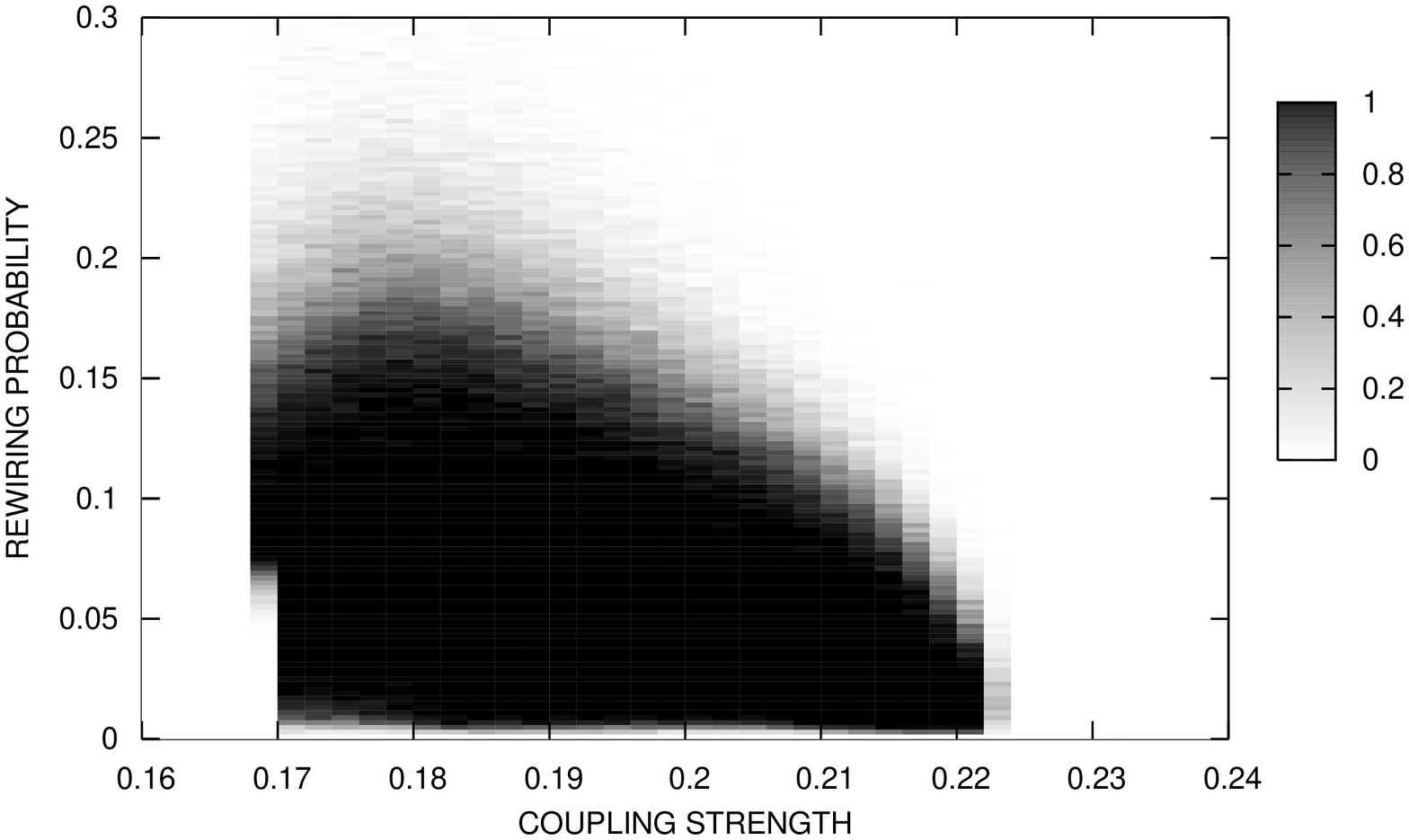}
\end{center}
\caption{Size of the basin of attraction of the synchronized cycles as
  a density plot in the space of coupling strength $\epsilon$ and
  random rewiring probability $p$, for a system of size $100$. Here
  transience is $10000$.}
\end{figure}

It is evident then that there is a {\em non-monotonic enhancement of
synchronization with increasing randomness in coupling connections},
and synchronization is most enhanced in some window of $p$ values.
This is in sharp contrast to the monotonic increase in synchronization
with increasing randomness in connectivity, in the strong coupling
regime in the {\em same} system. 

\begin{figure}[ht]
\label{fig4}
\begin{center}
\includegraphics[width=1\linewidth]{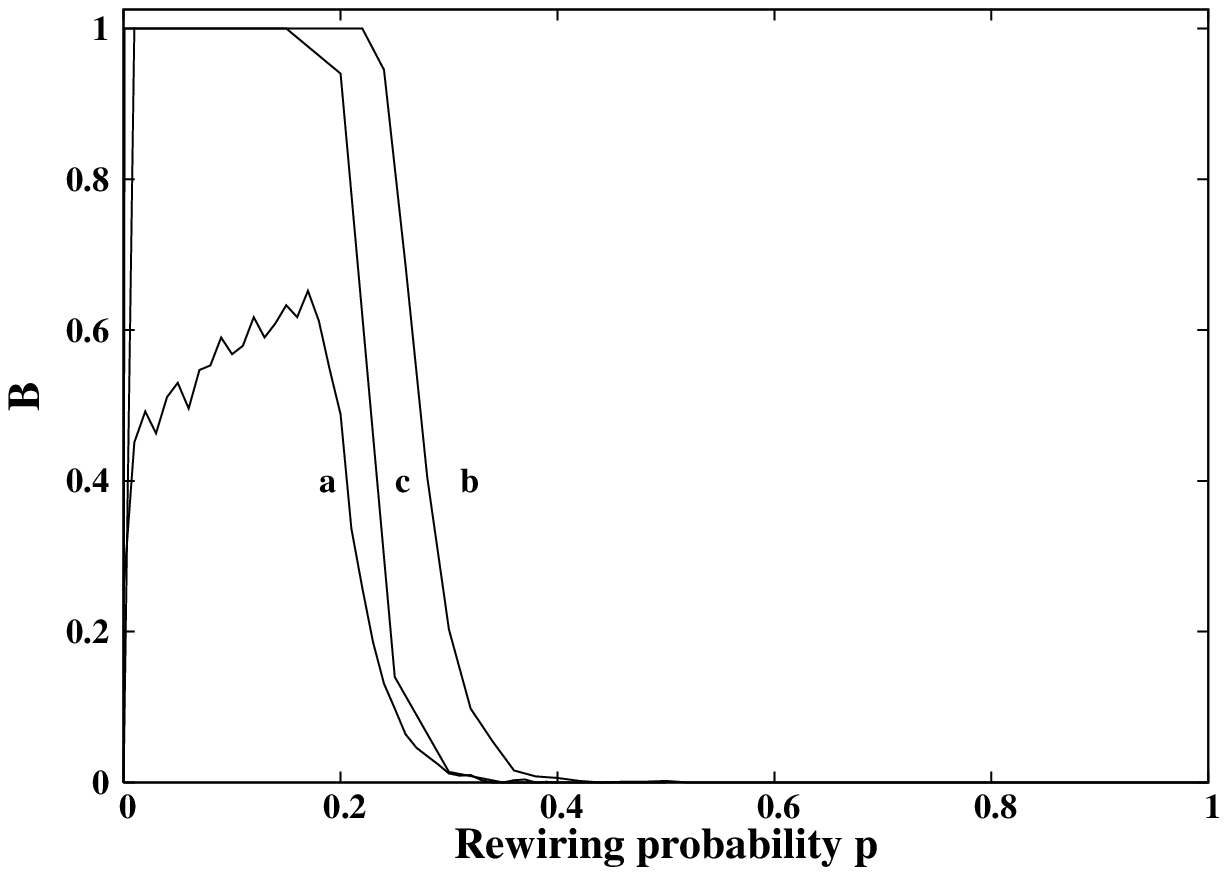}
\end{center}
%\mbox{\epsfig{file=l100pvsbasin.eps,width=10truecm,angle=270}}
%\mbox{\epsfig{file=Basin_vs_p_diff_eps.ps,width=10truecm,angle=270}}
\caption{Size of the basin of attraction of the synchronized state vs
  rewiring probability $p$ for a system of size $100$ for coupling
  strengths (a) $\epsilon=0.16$, (b) $\epsilon = 0.19$ (c)
  $\epsilon=0.20$. Here transience is $5 \times 10^5$.}
\end{figure}

\begin{figure}[ht]
\label{fig3}
\begin{center}
\includegraphics[height=1\linewidth]{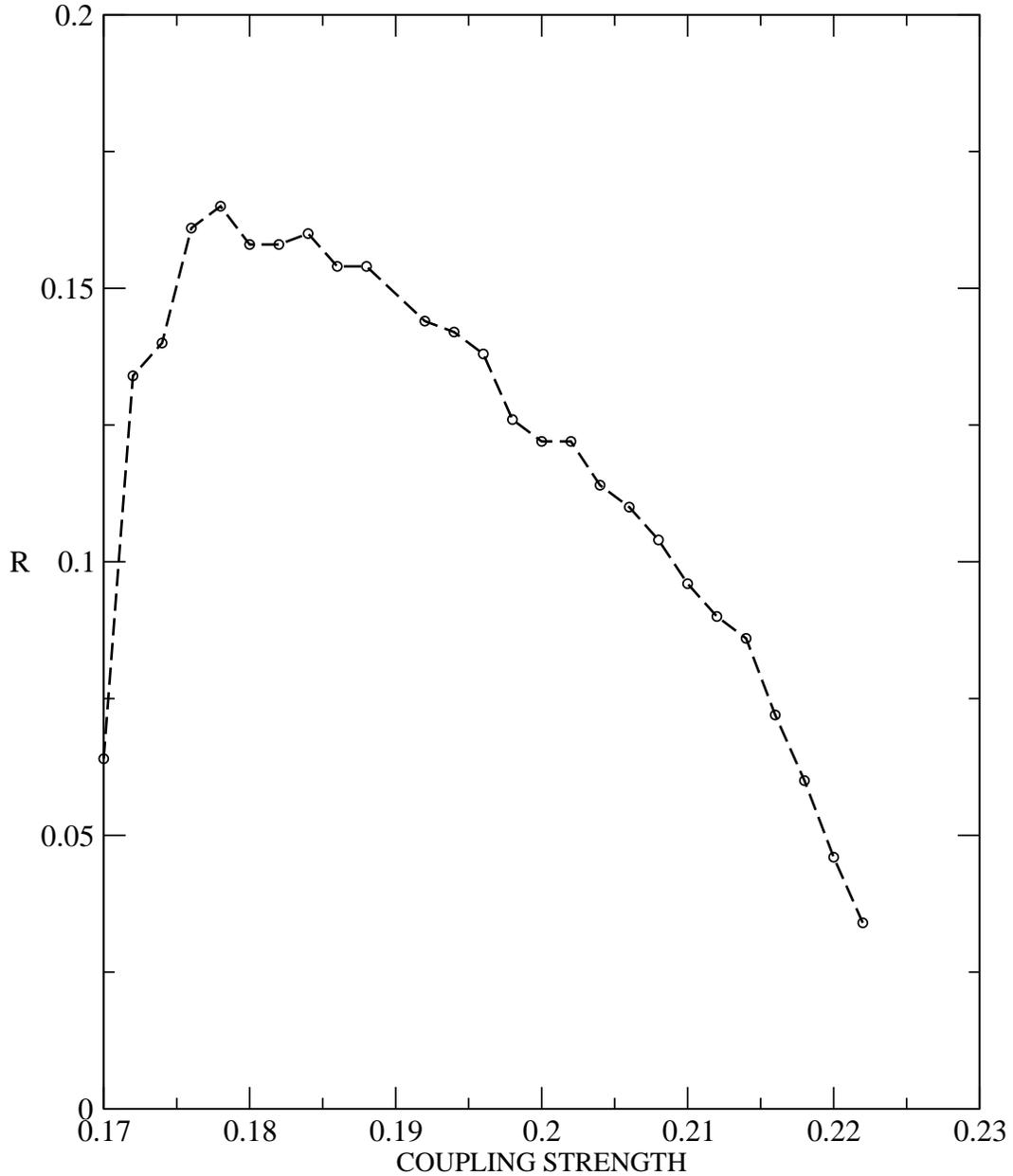}
\end{center}
\caption{Range of rewiring probability $R$, over which synchronization
  is achieved vs.  coupling strength $\epsilon$. Here system size is
  $100$ and transience is $10000$.}
\end{figure}

\section{Efficiency of synchronization}

We calculate the average time taken for systems with random initial
conditions to synchronize, and denote this as $<T>$. This quantity,
which is a measure of the {\em efficiency of synchronization}, is
displyed in Figs.~5-6.

It is evident from Fig.~5 that $<T>$ first decreases slightly as $p$
increases and then very sharply rises, by orders of magnitude,
especially for larger lattices. It is clear then that synchronization
is most efficient in some range of low $p$ values. So in order to
obtain fast and reliable synchronization it is a good strategy to
choose low $p$, as this {\em ensures a global attractor for the
  synchronized cycles, as well as a fast approach to the synchronized
  state from a random initial state.}

As a consequence of the above, the size of the basin of attraction for
the synchronized cycles, $B$, depends on the allowed transience.
Fig.~6 shows the increase in the basin $B$ as transience is increased.
However note that this increase is very slow, and increasing
transience by $2$ orders of magnitude changes the global attractor
width in $p$ space by a small amount.

\begin{figure}[ht]
\label{fig6}
\begin{center}
\includegraphics[width=1\linewidth]{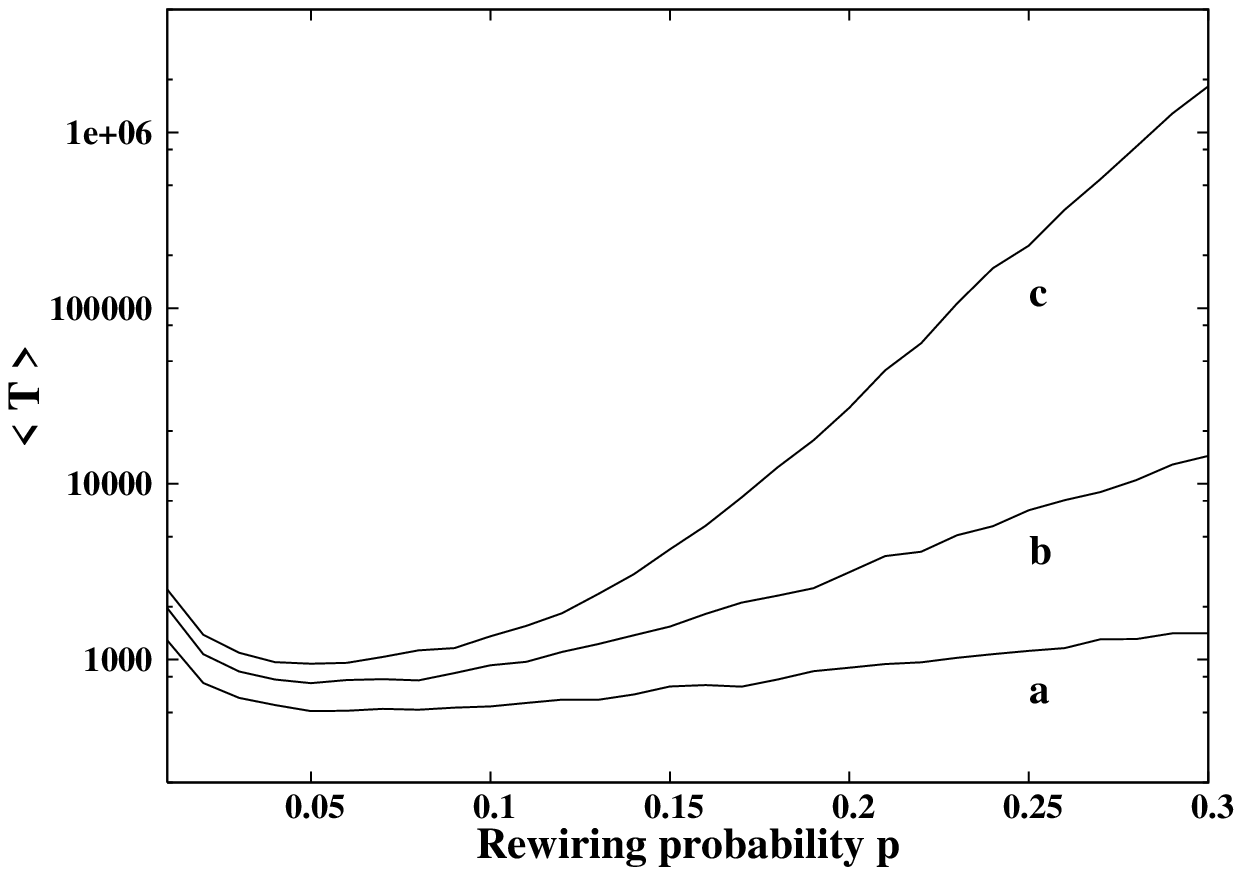}
\end{center}
\caption{Average time required to reach synchronization $<T>$ vs
  rewiring probabilty $p$, for coupling strength $\epsilon=0.19$ for
  system sizes $N = 25, 50,100$. The synchronized state is a global
  attractor for the system for the values of $p$ displayed here.}
\end{figure}

Caveat: it appears possible, from the above observations, that in the
limit of $T \rightarrow \infty$ a synchronized state may be reached
for larger lattices as well. But it should be underscored that
transience is always finite, and it is indeed very pertinent to know
what generically happens in the limit of finite, but large ($\sim O
(N^2)$), transience, as is the case for the studies in this paper.

\begin{figure}[ht]
\label{fig3}
\begin{center}
\includegraphics[width=1\linewidth]{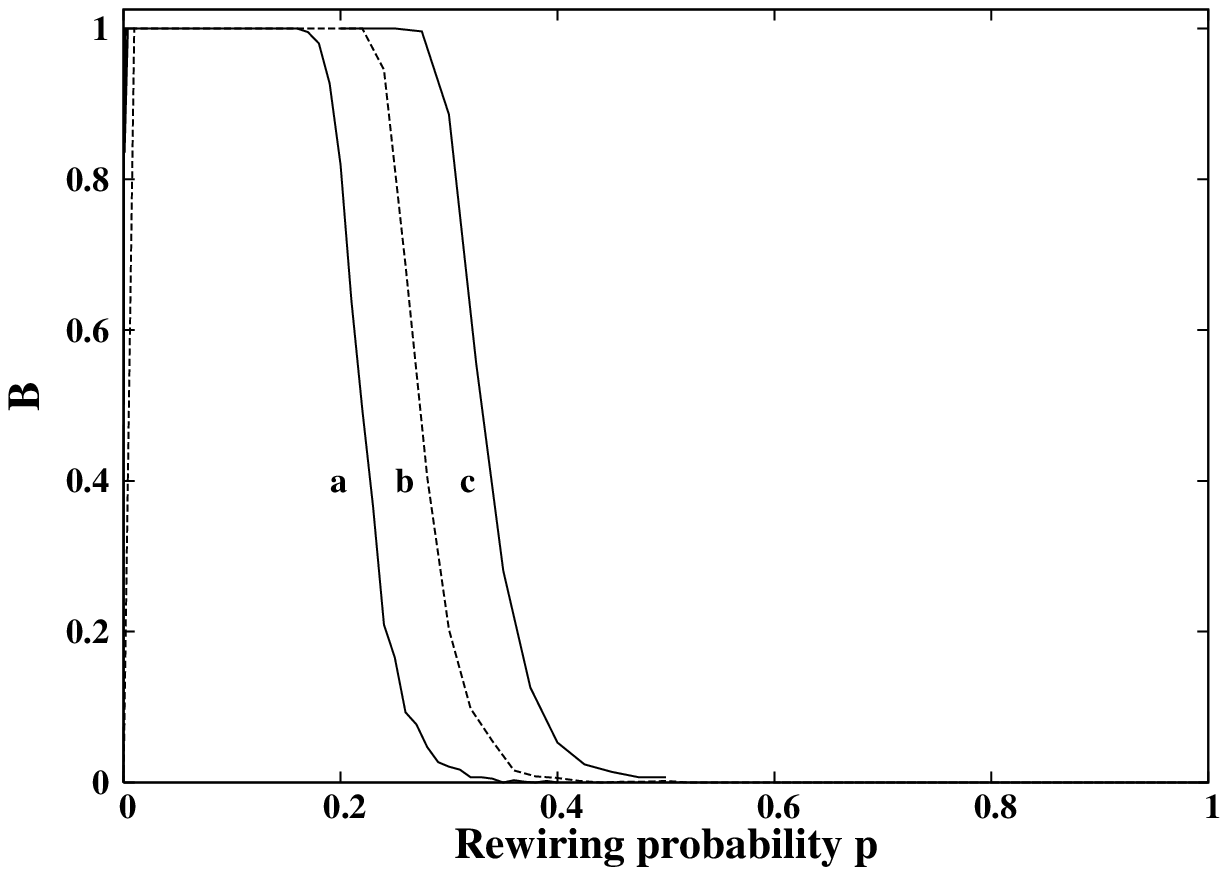}
\end{center}
\caption{Size of the basin of attraction of the synchronized state $B$
  vs rewiring probability $p$, for coupling strength $\epsilon$ for a
  system of size $100$ for maximum allowed transience equal to (a) $5
  \times 10^4$ (b) $5 \times 10^5$ (c) $5 \times 10^6$.}
\end{figure}

\section{Conclusions}

We have investigated the spatiotemporal dynamics of a network of
coupled chaotic logistic maps, with varying degrees of randomness in
coupling connections. We find a window in coupling parameter space, in
the weak coupling regime, where random rewiring induces spatiotemporal
order. Interestingly the basin of attraction for the synchronized
cycles varies non-monotonically with rewiring probability $p$.  As $p$
is increased, the basin of attraction of the synchronized state
rapidly increases. At an intermediate (small) value of $p$ the
synchronized state becomes the global attractor of the system with
(almost) all initial conditions being attracted to the synchronized
state. However, interestingly, as $p$ is further increased, the basin
of attraction of the synchronized state shrinks to zero again. Thus we
have strong evidence of the pronounced enhancement of spatiotemporal
order in an intermediate window of $p$. This is in marked contrast to
the monotonic increase in synchronization with increasing randomness
in connectivity, observed in the strong coupling regime of the same
system.

\end{document}